\documentclass[%
reprint,
superscriptaddress,
aps,
prb,
twoside,
twocolumn,
sort&compress,doi=false,
floatfix
]{revtex4-1}
\usepackage[utf8]{inputenc}
\usepackage[T1]{fontenc}
\usepackage{lmodern}

\usepackage{amsmath}    
\usepackage{graphicx}   
\usepackage{verbatim}   
\usepackage{color}      
\usepackage{subfigure}  
\usepackage{hyperref}   
\usepackage{xspace}
\usepackage[normalem]{ulem}
\usepackage{cancel}

\newcommand{\CN}{C$_3$N\xspace}
\newcommand{\Eformation}{E_\mathrm{form}}
\newcommand{\Ecoh}{E_\mathrm{coh}}
\newcommand{\muB}{\mu_\mathrm{B}}

\newcommand{\baslik}{Structural, electronic and magnetic properties of point defects in polyaniline (\CN) and graphene monolayers: A comparative study}

\begin{document}

\title{\baslik}
\author{K. Sevim}
\affiliation{Department of Physics, Izmir Institute of Technology, G{\"u}lbah\c ce Kamp{\"u}s{\"u}, 35430 Urla, Izmir, Turkey}
\author{H. Sevin\c cli}
\email{haldunsevincli@iyte.edu.tr}
\affiliation{Department of Materials Science and Engineering, Izmir Institute of Technology, G{\"u}lbah\c ce Kamp{\"u}s{\"u}, 35430 Urla, Izmir, Turkey}

\date{\today}

\begin{abstract}
The newly synthesized two-dimensional polyaniline (\CN) is structurally similar to graphene, and has interesting  electronic, magnetic, optical, and thermal properties. 
Motivated by the fact that point defects in graphene give rise to interesting features, like magnetization in an all carbon material, we perform density functional theory calculations to investigate vacancy and Stone-Wales type point defects in monolayer \CN.
We compare and contrast the structural, electronic and magnetic properties of these defects with those in graphene.
While monovacancies and Stone-Wales defects of \CN result in reconstructions similar to those in graphene, divacancies display dissimilar geometrical features.
Different from graphene, all vacancies in \CN have metallic character because of altered stoichiometry, those which have low-coordinated atoms have finite magnetic moments.
We further investigate the robustness of the reconstructed structures and the changes in the magnetic moments by applying tensile and compressive biaxial strain.
We find that, with the advantage of finite band gap, point defects in \CN are qualified as good candidates for future spintronics applications.
\end{abstract}

\maketitle

\section{Introduction}

Two-dimensional (2D) materials are promising candidates for the next-generation devices as they bring a wide range of unique properties.~\cite{novoselov,charlier,lee,rev1,rev2,rev3} 
An important advantage of these materials is the ease to modify their properties by nano-structuring schemes, some of which are creating heterostructures by stacking different types of monolayers, reducing their dimension to make ribbons or quantum dots, adsorbing functional groups to tailor their properties.~\cite{geim,ribbon,quantumdot,functional}
Defects are major ingredients that determine the material properties.
Defects and defect engineering in graphene related materials have been subjects of intense research during last years.~\cite{vicarelli:acsnano:2015,lusk:prl:2008}
Monovacancy (MV), divacancy (DV) and Stone-Wales (SW) defects are the most common defect types in graphene, where MV defect is a source of magnetism and lies at the heart of graphene spintronics studies.~\cite{han:nnano:2014,rostami,nanda,Valencia,ronchi:jphyschemc:2017,roche} 

Two-dimensional \CN (polyaniline) was first studied theoretically,~\cite{hu:pssb:2012,mizuno:syntheticmetals:1995}
and  was synthesized recently by Mahmood~\textit
{et al}.~\cite{mahmood} 
The atoms of C$_3$N monolayer are arranged in a honeycomb lattice structure like in graphene, and it is an indirect band-gap semiconductor\cite{mizuno:syntheticmetals:1995}.
It is synthesized by using direct pyrolysis of organic single crystal and fabricated by polymerization of 2,3-diaminophenazine.~\cite{mahmood,yangg}
Its electronic,~\cite{zhou:jmaterresearch:2017,xie:chemphys:2019} mechanical\cite{chen,zhou:jmaterresearch:2017}, thermal \cite{kumar,hong}, and magnetic \cite{Bafekry:AppliedSurfaceScience:2020,Bafekry:Carbon:2020}properties have been studied  and some possible applications were suggested such as anode material for batteries\cite{pathak}, photocatalytic\cite{singh} and nanosensor applications.~\cite{li,cui,makaremi,Bafekry:ChemicalPhysics:2019}
In addition, its zigzag nanoribbons have also been studied.~\cite{jalili,bagheri,Qingfang,XIA2018,ma2}

In the present study, we perform first-principles calculations  to investigate the structural, electronic and magnetic properties of  MV, DV, and SW defects on monolayer C$_3$N. 
We compare and contrast the structural, electronic and magnetic properties of the \CN with graphene.
Since it is a newly synthesized material, the properties of point defects in \CN are not studied experimentally yet and computational investigations are expected to guide future experimental observations. 

We show that, while having similar structural features, \CN-MV structures give rise to partially filled bands and have finite magnetic moments, whereas \CN-DV can structurally be distinguished from their graphene counterparts and induce magnetic moments unlike in graphene.
We also investigate the effects of tensile and compressive biaxial strain on defected polyaniline.

\section{Computational Methods}
	Points defects are modeled using the cluster approach or periodic boundary conditions, both of which bring in certain spurious effects. The main shortcoming of the cluster approach is the finite size effect and interaction with the edge states. Using periodic boundary conditions, one prevents edge states, but includes inter-defect couplings. Using large enough super cells, inter-defect coupling can be minimized, which is the method we choose. 
	We note that, some quantitative features, such as the defect induced band gap value, or the exact value of the magnetic moment 
	can be affected from the choice of the super cell size, which also determines the defect density.
	Still, the qualitative features should be independent of the super cell size, provided that it is large enough.
	We also note that, electronic band diagrams, as well as the density of states plots, should be interpreted in accordance with the choice of super cell size.

The calculations are performed within the framework of density functional theory (DFT) as implemented in the \verb|VASP| code.~\cite{kresse1,kresse2}
The projector augmented wave (PAW) potentials are used with the Perdew-Burke-Ernzerhof (PBE) functionals of the generalized gradient approximation (GGA).~\cite{perdew1,perdew2}
Plane wave energy cutoff values of 400~eV and 500~eV are used for graphene and C$_{3}$N, respectively.
The structures are relaxed until the force on each atom is less than $0.001$~eV/\AA, and the self consistency tolerance is set to $10^{-6}$~eV. 
Super cells consisting of 128 atoms in their pristine structures, which correspond to  8$\times$8$\times$1 and 4$\times$4$\times$1 super cells for graphene and C$_3$N, are used so that the interaction between defects is negligible. 
The vacuum spacing between layers is set to 10~\AA.  
During the DFT calculations, the k-point sampling is carefully examined to ensure that the calculation results are converged. 
For defective super cells, reciprocal space was sampled with 2$\times$2$\times$1 k-points using the Monkhorst-Pack scheme.~\cite{monkhorst} 
The k-point grid is chosen between 5$\times$5$\times$1 and 9$\times$9$\times$1 to calculate the density of states (DOS).

In order to investigate the stability of defective structures at elevated temperatures, we perform \textit{ab initio} molecular dynamics (AIMD) simulations. The canonical ensemble with a Nos\'{e}-Hoover thermostat are used at 500 K. The time step used in AIMD simulations is 1 fs, and the total duration 10 ps(10000 steps). 

\begin{figure}[t]
	\centering
	\includegraphics[scale=0.12]{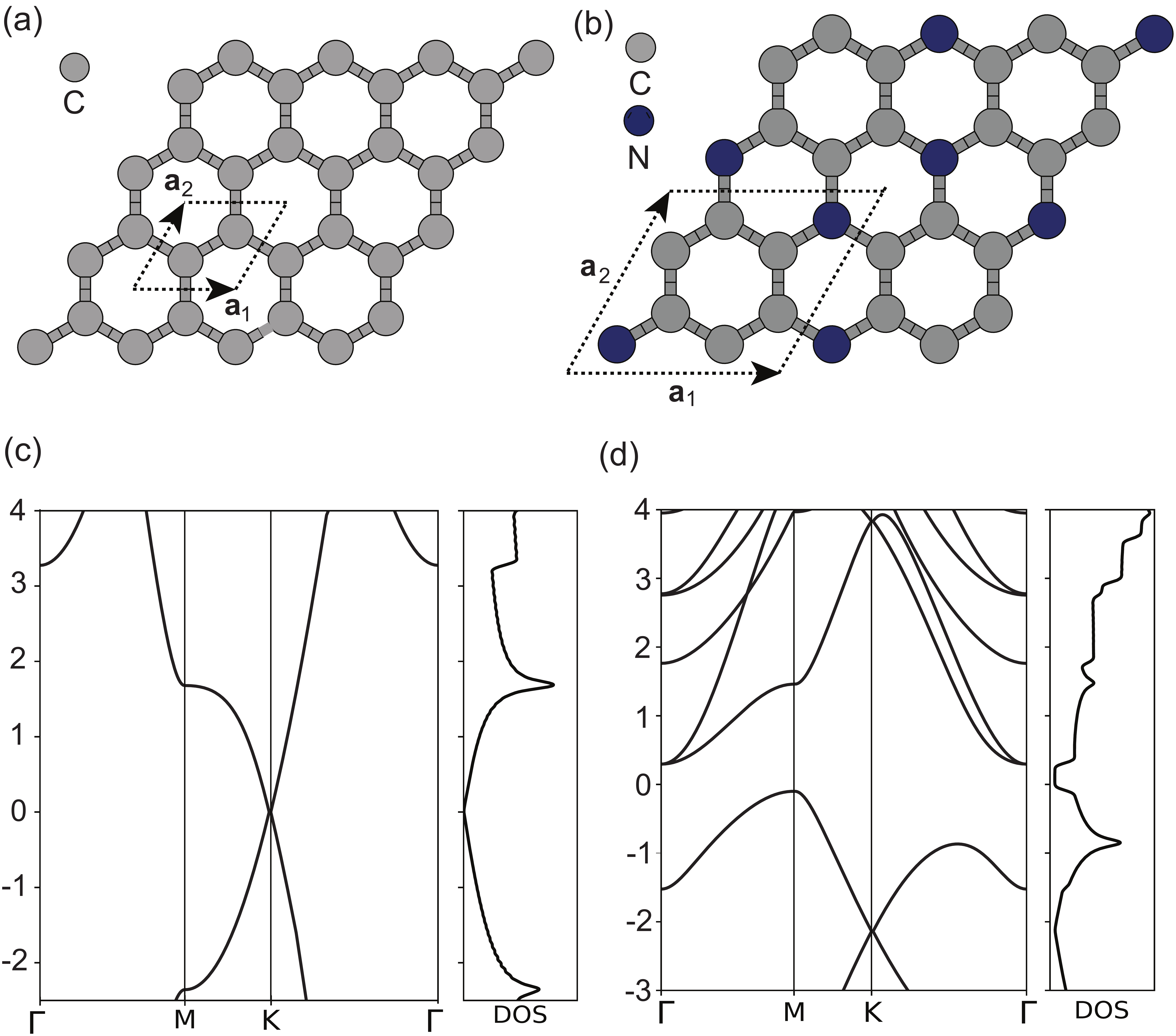}
	\caption{\label{fig:fig1} Honeycomb structures of pristine graphene and \CN monolayers are shown in (a) and (b), respectively. The unit cells are depicted with dashed lines. Electronic bands and DOS of the pristine graphene and \CN are plotted in (c) and (d), respectively, where the zero of the energy is set to the Fermi level.
	}
\end{figure}

The cohesive energy $\Ecoh$ is defined as 
\begin{eqnarray}
	\Ecoh &=& (n_\mathrm{C} E_\mathrm{C} + n_\mathrm{N} E_\mathrm{N} -E_\mathrm{tot})/(n_\mathrm{C} + n_\mathrm{N}) \label{coh},
\end{eqnarray}
where  $n_\mathrm{C}$ and $n_\mathrm{N}$ denote the number of C and N atoms, $E_\mathrm{C}$ and $E_\mathrm{N}$ are energies of single C and N atoms, $E_\mathrm{tot}$ is the total energy of the structure under consideration.
Defect formation energy is obtained as
\begin{eqnarray}
	\label{form}
	\Eformation &=& E_\mathrm{tot}^\mathrm{defective}+\sum_{i = \mathrm{C,N}} n'_i \mu_i-E_\mathrm{tot}^\mathrm{pristine},
\end{eqnarray}
with $E_\mathrm{tot}^\mathrm{pristine}$ and $E_\mathrm{tot}^\mathrm{defective}$ being total energies of pristine and defective super cells, respectively. 
The number of removed atoms from species $i$ is denoted with $n'_i$.
The chemical potentials
$\mu_\mathrm{C}{=}E_\mathrm{tot}^\mathrm{Graphene}/n_\mathrm{tot}$ and $\mu_\mathrm{N}{=}(E_\mathrm{tot}^\mathrm{C_3N}-n_\mathrm{C}\mu_\mathrm{C})/n_\mathrm{tot}$ are those of single C and N atoms of the pristine structures. 
The magnetic stabilization energy is calculated using
$\Delta E_\mathrm{mag} = E_\mathrm{nonmag}-E_\mathrm{mag}$,
where  $E_\mathrm{nonmag}$ is the total energy of spin unpolarized calculation, and $E_\mathrm{mag}$ it the total energy of the spin polarized calculation.

\section{Results and Discussion}
\subsection{Structural Properties}
The pristine graphene and \CN monolayers and their unit cells can be seen in Fig.~\ref{fig:fig1}. The lattice parameters are 2.47~{\AA} and 4.86~{\AA} for graphene and \CN, respectively.~\cite{zhou:jmaterresearch:2017}
In C$_{3}$N monolayer, both C-N and C-C bond lengths are approximately 1.40~{\AA}, with C-N bond being slightly longer, which is in good agreement with previous theoretical results 1.404~{\AA} and 1.403~{\AA}\cite{Bafekry:PCCP:2019}.
The C-C bond length is 1.42~{\AA} in graphene, for comparison. 
We consider three defect types, MV, DV and SW defects.
For graphene, these correspond to three structures, whereas for \CN there are two possible structures for each defect.
The MV can be of C or N type, \CN-MV(C) and \CN-MV(N). 
Since DV and SW involve a pair of atoms, they can be (CC) or (CN) type.
Namely, we have six defective structures of \CN.


\begin{figure}[b]
	\centering
	\includegraphics[scale=0.4]{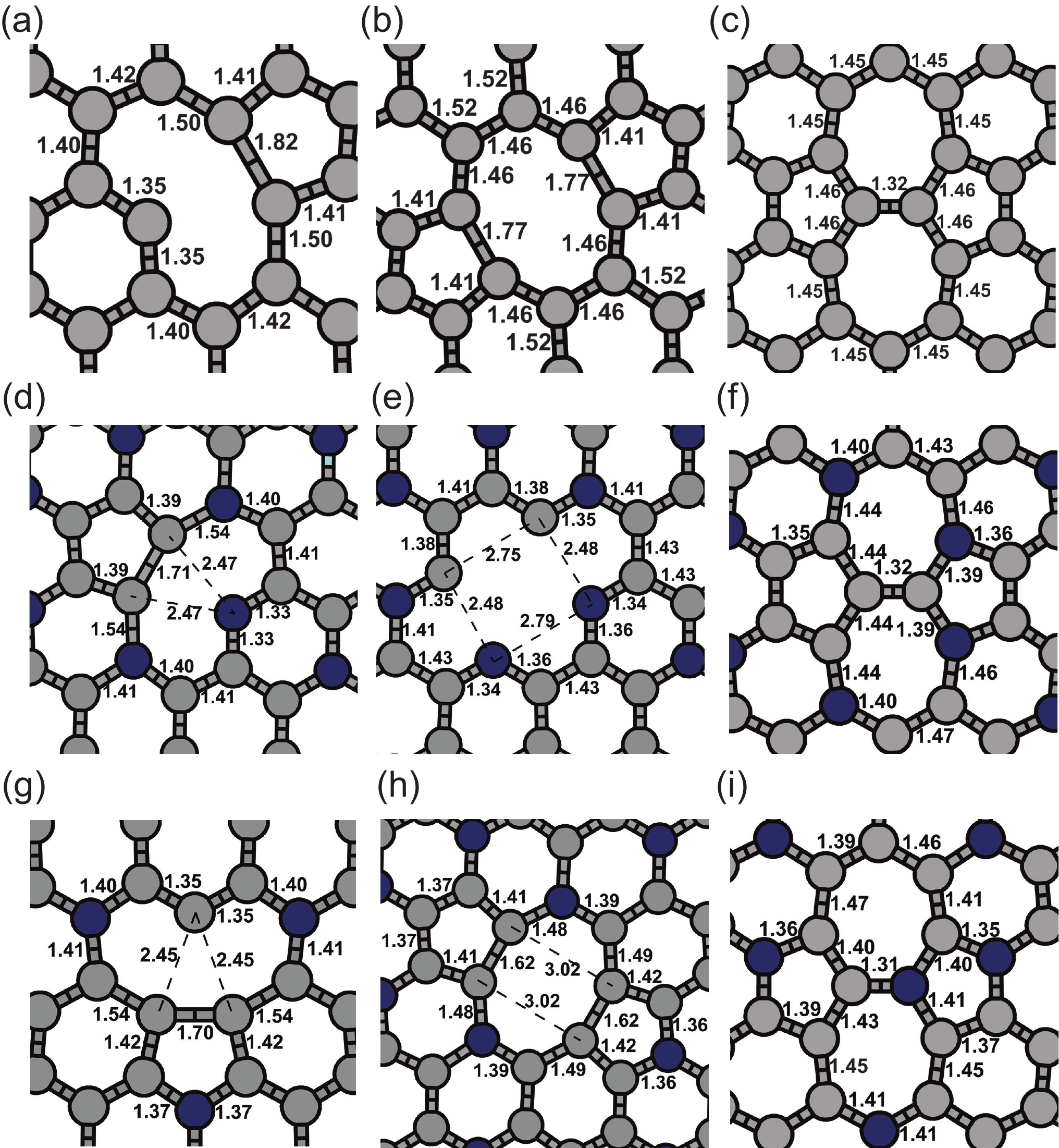}
	\caption{\label{fig:yapisal} Graphene and C$_{3}$N defect lattice structures (a) graphene MV(gr-MV) (b) graphene DV(gr-DV) (c) graphene Stone-Wales(gr-SW) (d) C$_{3}$N-MV(C) (e) C$_{3}$N-DV(CC) (f) C$_{3}$N-SW(CC) (g) C$_{3}$N-MV(N) (h) C$_{3}$N-DV(CN) (i) C$_{3}$N-SW(CN)}
\end{figure}

We first investigate the structural aspects of gr-MV defect.
Removing a carbon atom from the hexagonal lattice, the remaining structure is three-fold symmetric before relaxation.
Three carbon atoms belonging to the opposite sublattice with respect to the removed atom are two-coordinated.
If these atoms are moved towards the vacancy site without breaking the three-fold symmetry, they would have four bonds each, which is  not favorable within sp$^2$ hybridization.
Rather, there are three possible states which are more favorable. 
This is achieved by breaking the three-fold symmetry and forming a pentagon-nonagon pair (5-9 reconstruction) having only one two-coordinated carbon, which is known as the Jahn-Teller distortion.~(Fig.~\ref{fig:yapisal}a)
However, it is not trivial to obtain this structure in the simulations, which is the reason for having two different structures reported in the literature\cite{gr-MV-3}. 
The structure needs to be perturbed to break the symmetry and the super cell should be large enough to distribute the local strain over the neighborhood.
If these two conditions are not fulfilled, the simulation results in a metastable state with a three-fold symmetric geometry and three two-coordinated atoms, or even if the symmetry is slightly broken there is no new bond formation.
Once the conditions are fulfilled, the structure finds the stable geometry, where two of the low-coordinated carbon atoms form a 1.82~{\AA} long bond to stabilize. 
The remaining two-coordinated carbon shortens its bonds down to 1.35~{\AA}.
The formation and cohesive energies for gr-MV are 7.89~eV and 7.91~eV, respectively (see Table~\ref{tab:formation})
The difference in total energies between the stable and metastable structures is 23~meV.
The magnetic properties of gr-MV are closely related with this reconstruction, which are discussed in the next section.

In \CN-MV(C) (Fig.~\ref{fig:yapisal}d), there are two nitrogen and one carbon atoms which are two-coordinated, therefore there is no three-fold symmetry as in gr-MV.
Two-coordinated carbon atoms form a new bond leaving the nitrogen two-coordinated, and the 5-9 reconstruction is achieved without any need to perturb the geometry.
The length of the newly formed bond is 1.71~{\AA} in this case.
The situation is similar for \CN-MV(N) (Fig.~\ref{fig:yapisal}g). 
This time, the initial geometry fulfills three-fold symmetry, like in graphene.
Perturbing the system, the symmetry is broken and a pentagon-nonagon structure is formed.
The bond lengths of the reconstructed \CN-MV(N) are very close to those in \CN-MV(C).
The defect formation energies for \CN-MV(C) and \CN-MV(N) are 4.71~eV and 5.84~eV, respectively. The cohesive energies are close, 7.04~eV and 7.06~eV, which agrees well with previously reported values of 6.98~eV and 7.02~eV, respectively \cite{Guo:AppliedSurfaceScience:2019}. Recently, Xie \textit{et al.} have obtained different defect formation energy values. Since they have $32$ atoms in $2 \times 2$ supercell of \CN, their formation energies larger than our values.\cite{Xie:ChemicalPhysics:2019}
The reconstructed structures of \CN-MV that we report are different from those in the literature.
Ma \textit{et al.} did not obtain any 5-9 formations for \CN-MV(C) or \CN-MV(N).~\cite{ma}
Rather, they have three two-coordinated atoms at the defect site, which we show to be metastable states and have differences in electronic and magnetic properties, when compared to the ground states shown here. These properties will be discussed in the next section.

\begin{figure*}
	\centering
	\includegraphics[width=0.28\textwidth]{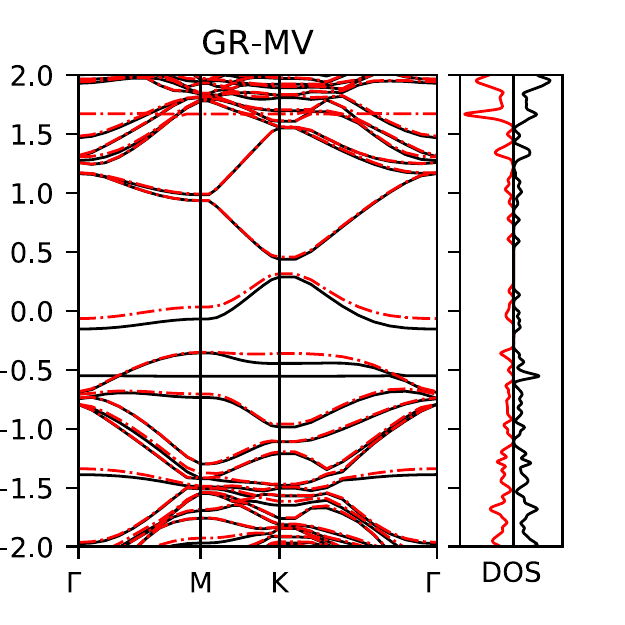}\quad\hspace{-0.5cm}
	\includegraphics[width=0.28\textwidth]{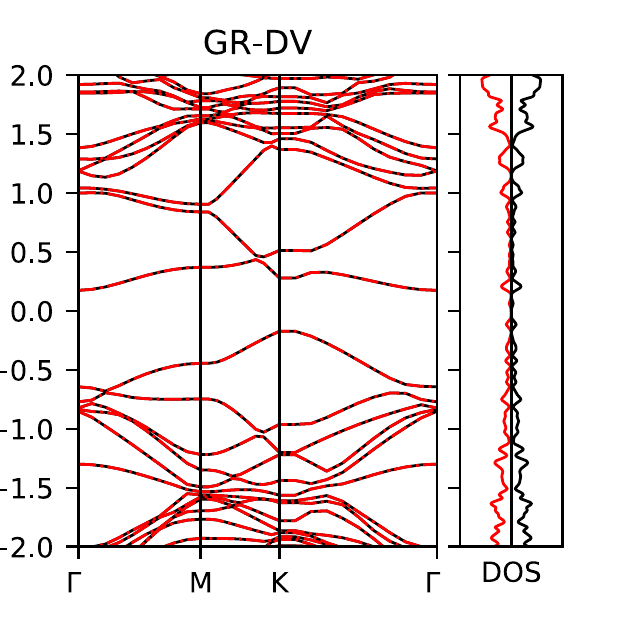}\quad\hspace{-0.5cm}
	\includegraphics[width=0.28\textwidth]{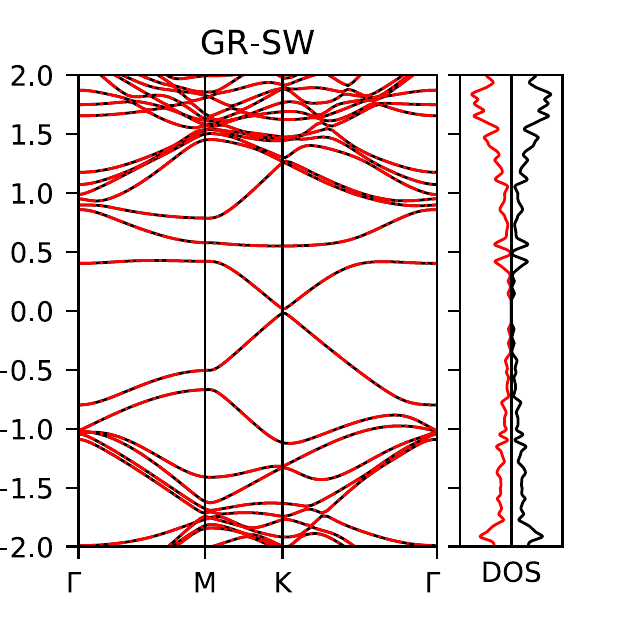}\hspace{5cm}
	\medskip
	\includegraphics[width=0.28\textwidth]{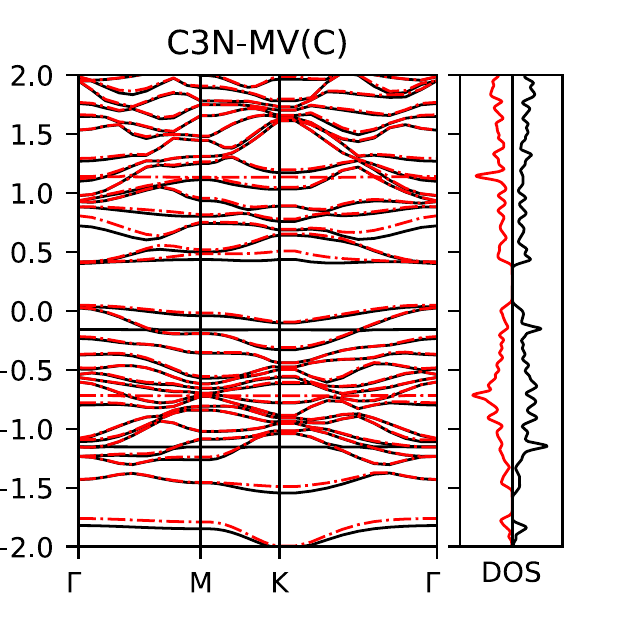}\quad\hspace{-0.5cm}
	\includegraphics[width=0.28\textwidth]{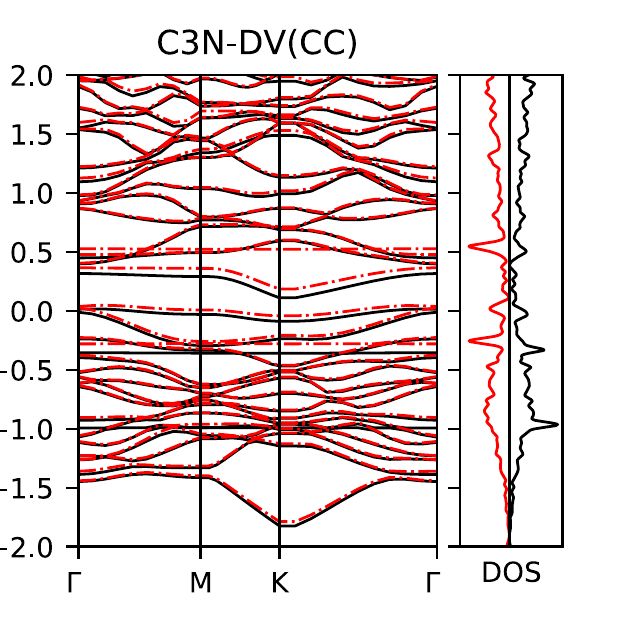}\quad\hspace{-0.5cm}
	\includegraphics[width=0.28\textwidth]{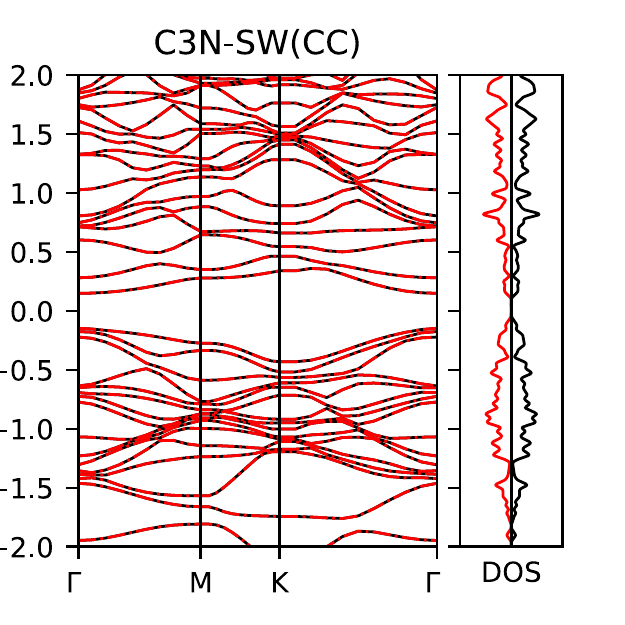}\hspace{5cm}
	\medskip
	\includegraphics[width=0.28\textwidth]{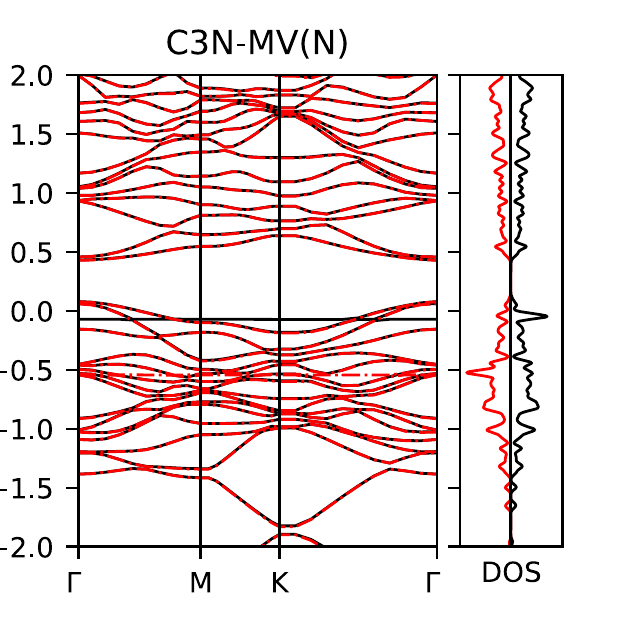}\quad\hspace{-0.5cm}
	\includegraphics[width=0.28\textwidth]{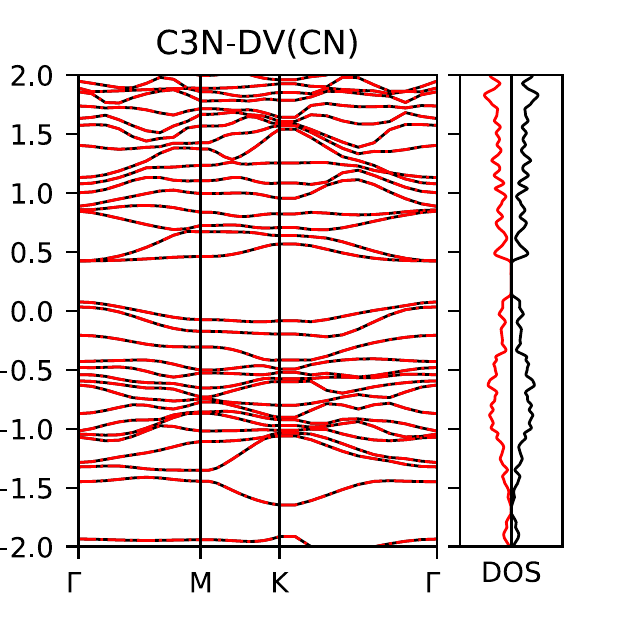}\quad\hspace{-0.5cm}
	\includegraphics[width=0.28\textwidth]{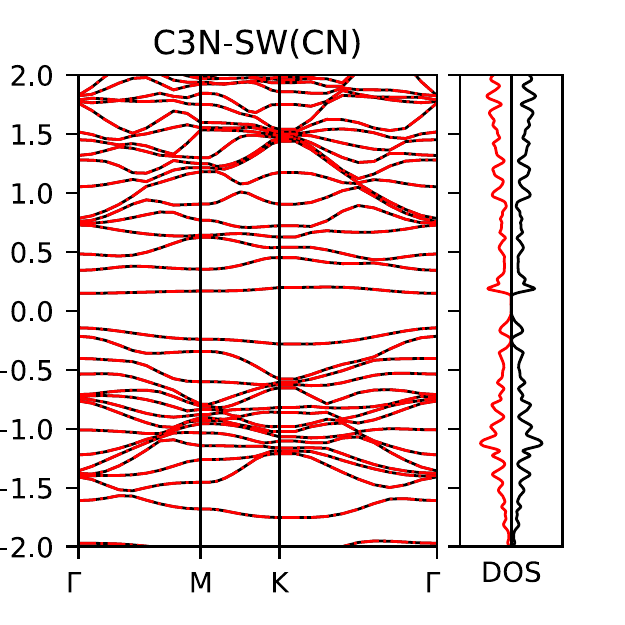}
	\caption{Spin polarized band structure and DOS graphs of defective graphene and C$_{3}$N structures. Black-solid (red-dashed) curves represent spin-up (spin-down) electrons. The first row contains graphene structures, the remaing two are \CN structures. Each column represents a defect type, \textit{i.e.} monovacancy, divacancy and Stone-Wales defects. The zero of the energy is set to the Fermi level.
	}	
	\label{fig:band-dos}
\end{figure*}


\begin{table}[t]
	\caption{\label{tab:formation}Defect formation energy $\Eformation$, cohesive energy $\Ecoh$, magnetic moment $m$, and magnetic stabilization energy $\Delta E_\mathrm{mag}$ values for different defect types. Cohesive energies of the pristine structures of graphene and \CN are 7.98~eV and 7.08~eV for comparison.}
	\centering
	\begin{tabular}{lccccc}
		\hline\hline
		Structure &  & $\Eformation$(eV) & $\Ecoh$(eV) & m($\mu_\mathrm{B}$) & $\Delta E_\mathrm{mag}$(eV) \\
		\hline
		gr-MV &  & 7.89 & 7.91 & 1.15 & 0.09\\
		gr-DV &  & 7.56 & 7.92 & - & -\\
		gr-SW &  & 5.06 & 7.94 & - & -\\
		C$_{3}$N-MV(C) \rule{0pt}{4ex} &  & 4.71 & 7.04 & 0.99 & 0.18\\
		C$_{3}$N-MV(N) &  & 5.84 & 7.06 & 0.29 & 0.72\\
		C$_{3}$N-DV(CC) \rule{0pt}{4ex} &  & 4.90 & 7.03 & 1.23 & 0.16\\
		C$_{3}$N-DV(CN) &  & 6.32 & 7.05 & - & -\\
		C$_{3}$N-SW(CC) \rule{0pt}{4ex} &  & 5.01 & 7.04 & - & -\\
		C$_{3}$N-SW(CN) &  & 3.42 & 7.06 & - & -\\
		\hline\hline
	\end{tabular}
\end{table}

Divacancy structure in graphene, gr-DV, is obtained by removing a pair of neighboring carbon atoms. 
The gr-DV has a more stable structure than the gr-MV because all atoms are three-coordinated after the reconstruction. 

The hexagons which lost one of their corners close to form pentagons, and a pentagon-octagon-pentagon (5-8-5) geometry is obtained (Fig.~\ref{fig:yapisal}b).~\cite{ciraci,gr-DV-1}

Newly formed bonds are 1.77~{\AA} in length, similar to the bonds in gr-MV.
$\Eformation$ is 7.56~eV and lesser than that of gr-MV, while $\Ecoh$ is 7.92~eV and higher than that of gr-MV. (see Table~\ref{tab:formation})
In \CN, there are two possibile divacancy formations.
One is removal of a carbon-carbon pair, \CN-DV(CC), the other is removal of carbon-nitrogen pair, \CN-DV(CN), as shown in Figures~\ref{fig:yapisal}e and \ref{fig:yapisal}h, respectively.
\CN-DV(CN) undergoes the 5-8-5 reconstruction just like in graphene but this is not the case for \CN-DV(CC). 
The open hexagons do not form pentagons, and a hole with fourteen edges and four two-coordinated atoms (two carbon, two nitrogen) are left.
We have performed lattice relaxation under biaxial compressive strain and find that even though the openings of the hexagons tend to close, there is no bonding and the 5-8-5 reconstruction does not take place even at a strain value of -0.06.
Nevertheless, $\Eformation$ for \CN-DV(CC) is 4.90~eV, and less than \CN-DV(CN),
and $\Ecoh$ of \CN-DV(CN) is larger that of \CN-DV(CC), as expected.


Stone-Wales (SW) defect is formed by rotating a bond by 90$^\mathrm{o}$.
The number of atoms is preserved for each species.
Two possible SW defects, in polyaniline are \CN-SW(CC) and \CN-SW(CN).
After relaxation two pentagons and two heptagons form a 5-7-7-5 geometry for both structures considered (see Fig.~\ref{fig:yapisal}(c,f,i)).
All atoms are three-coordinated and the rotated bond have similar lengths around 1.32~{\AA} for all structures.
Unlike the DV defect, the structural aspects of the SW defect in \CN is not sensitive to the pair type (C-C or C-N).
On the other hand \CN-SW(CN) has considerably lower energy than \CN-SW(CC), by about 1.59~eV. Accordingly, its $\Ecoh$ is larger and $\Eformation$ is lower.

Comparing the $\Eformation$ and $\Ecoh$ of graphene defects, gr-SW is the most stable one. The gr-MV is the least stable because of the two-coordinated carbon atom. 
Similarly, the SW defects are among the most stable ones in \CN, as their $\Eformation$ are low, but there is no clear distinction between different defect types in terms of $\Eformation$ or $\Ecoh$. 
\CN-DV(CC) has the lowest $\Ecoh$ as it has four two-coordinated atoms.
\CN-MV(N) and \CN-SW(CN) have the highest $\Ecoh$,
whereas \CN-MV(N) and \CN-DV(CN) have the highest $\Eformation$.
Namely, $\Ecoh$ and $\Eformation$ are not inversely related like in graphene defects, which is interpreted as a consequence of charge redistribution and unequal strain energies stored around the defect sites.

\begin{figure}[t]
	\centering
	\includegraphics[scale=0.28]{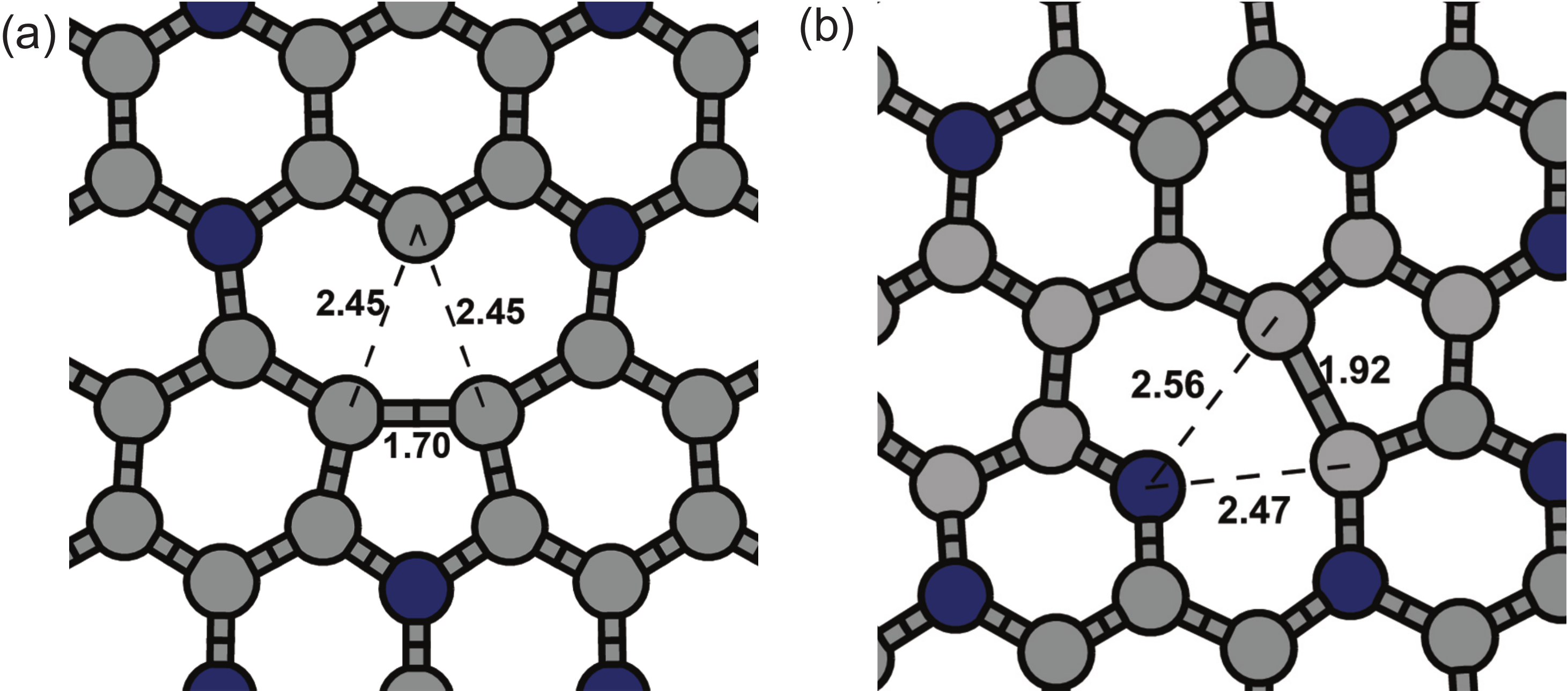}
	\caption{\label{fig:mnv-mig} Defect migration in \CN-MV(N) after the AIMD simulation. (a) \CN-MV(N) initial structure (b) \CN-MV(N) final structure.} 
\end{figure}

We test all the six defective structures using AIMD to be sure that they are stable at finite temperatures. We observe appreciable out-of-plane deformation around the defect atoms, which was absent when optimized at zero temperature. In \CN-SW(CC) and \CN-SW(CN) structures, the deformation is more pronounced than the other structures. 
We observe that there is no dissociation in any of the defective structures. All defects are stable at 500~K except \CN-MV(N), which is found to be metastable at this temperature. We remind that \CN-MV(N) is found to be dynamically stable according to zero temperature phonon calculations. 
We observe a migration of defect in \CN-MV(N) structure, for which the initial and final geometries are shown in Fig~\ref{fig:mnv-mig}. 
At the very first steps of the AIMD simulation, two carbon atoms from the pentagon migrate to convert the nonagon to two regular hexagons, leaving the nitrogen of the pentagon two-coordinated.
This final structure remains unchanged until the end of the simulation.
The total energy difference between the initial and final geometries is found to be $1.28$~eV. 
The migration can not be observed during zero tempeture structural relaxation process. 
Besides the presence of a two-coordinated nitrogen atom, the stabilized \CN-MV(N) structure has zero magnetic moment.
This is interpreted as being due to pairing of the dangling bond with the extra electron of the nitrogen.

\subsection{Electronic and Magnetic Properties}
\label{sec:electronic_magnetic}

\begin{figure}[t]
	\centering
	\includegraphics[scale=0.37]{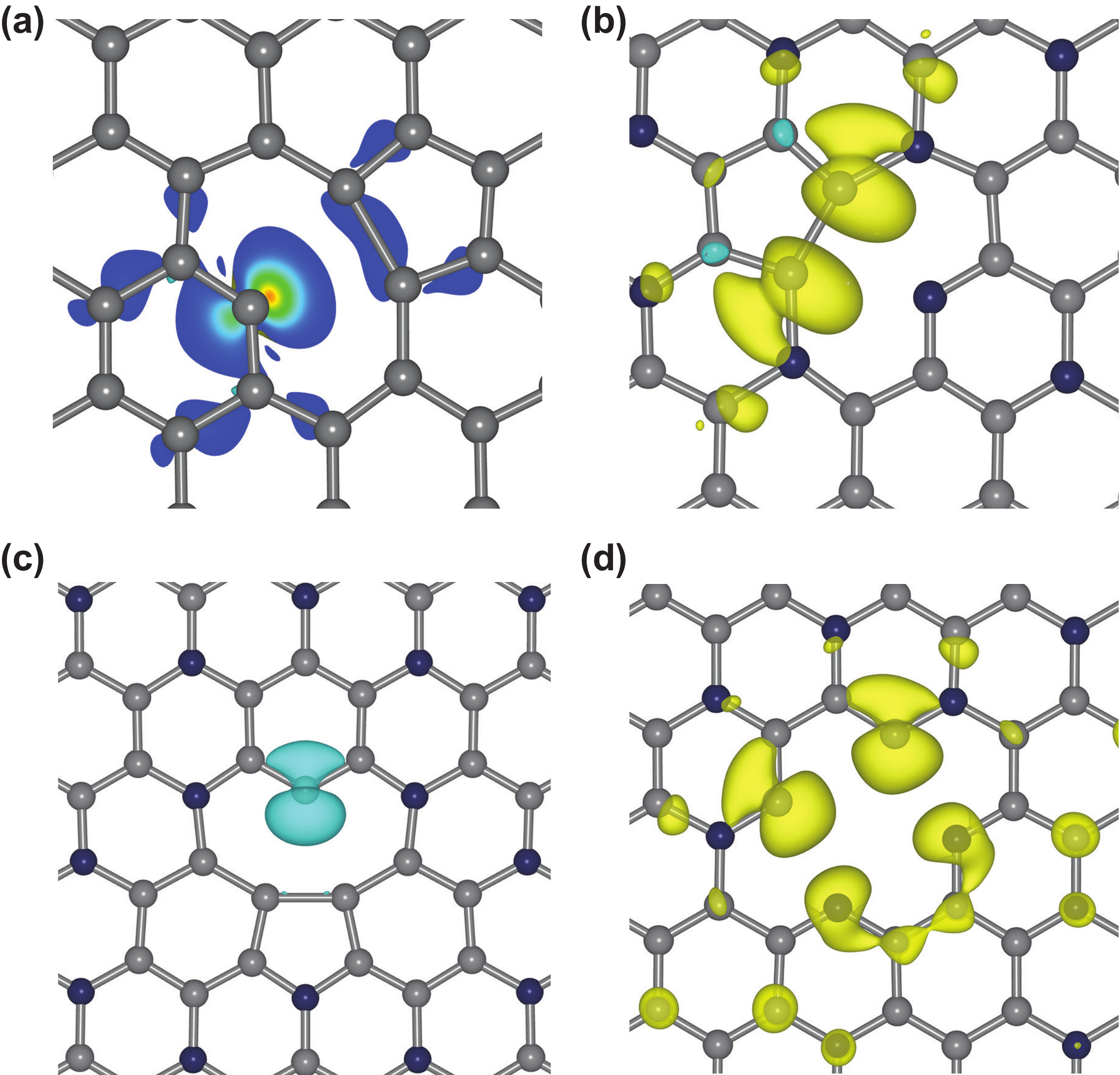}
	\caption{\label{fig:chgu-d} Magnetization of graphene and C$_{3}$N defcts. (a) gr-MV (b) C$_{3}$N MV(C) (c) C$_{3}$N MV(N) (d) C$_{3}$N DV(CC), the isovalue for the spin density plots ($\Delta S=\rho_{\uparrow}-\rho_{\downarrow}$) is 0.0025 e/au$^3$ (positive values are in yellow, negative values are in blue).} 
\end{figure}


Graphene is a zero band gap semiconductor, with its valence ($\pi$) and conduction ($\pi^*$) bands having linear dispersions touching at the corners of the Brillouin zone and forming Dirac cones (Fig~\ref{fig:fig1}c).~\cite{wallace:physrev:1947}
On the other hand, \CN has an indirect band gap, with its valence band maximum at M and conduction band minimum at $\Gamma$ (Fig.~\ref{fig:fig1}d)
The band gap is predicted as 0.40~eV according to PBE calculations,~\cite{makaremi,kumar} and the corrected band gap is 1.04~eV when  HSE06 hybrid functionals are used.~\cite{hse06,zhang,makaremi}
Corresponding band structures and DOS for defective graphene and \CN, as obtained from spin polarized DFT calculations are plotted in Fig.~\ref{fig:band-dos}.

In graphene, the $\pi$- and $\pi^*$-bands develop a finite energy gap when defects are introduced (see the first row of Fig.~\ref{fig:band-dos}). Fermi energy lies within the gap for gr-DV and gr-SW, but it crosses the $\pi$-band in gr-MV.
Moreover, there exist a flat band below the Fermi level at -0.5~eV belonging to the majority spin, and another belonging to the minority spin at around 1.7~eV.
Spin degeneracy is preserved and no flat bands occur in gr-DV and gr-SW structures. These findings are in agreement with the literature.~\cite{ronchi:jphyschemc:2017,Valencia,DV,SW}

Nitrogen doped graphene could be viewed as an intermediate state between graphene and \CN.
Structurally, the bond lengths of the dopant atom were reported to be 1.40~{\AA} and almost no distortion in the planar structure of graphene was observed.~\cite{fan:rscadv:2013,rani:rscadv:2013}
A band gap, whose value depends on the super cell size, is introduced at the Dirac point due to breaking of the sublattice symmetry.~\cite{casolo:jphycemc:2011}
Fermi level is shifted to higher energies and no magnetic moment is found, that is the extra electron is accounted for charge doping.~\cite{rani:rscadv:2013,pizzochero:jpcm:2015}
In \CN unit cell there are even number of electrons and the Fermi level is inside the band gap.
Vacancies, however, alter the number of electrons in the system and electronic structures of defective systems differ, as it will be discussed below.

The gr-MV structure is arguably the most studied defect in graphene.~\cite{gr-MV-3,gr-MV-2,ronchi:jphyschemc:2017, Valencia,Zhang2016,Hashimoto2004,Ugeda2010,ElBarbary2003,Lehtinen2004,Yazyev2007,Miranda2016,Casartelli2013}
The main reason for this is the magnetization in a pure carbon system owing to the dangling bonds around the defect sites.
The reason of magnetization can be understood referring to Lieb's theorem on bipartite lattices.~\cite{Lieb}
At the dilute limit, the magnetic moment is expected to converge to 2~$\muB$, 1~$\muB$ originating from the $\sigma$- and 1~$\muB$ from the $\pi$-bonds.
Indeed, computations with very large super cells~\cite{rodrigo:carbon:2016}
and those employing  hybrid functional methods~\cite{ronchi:jphyschemc:2017,Valencia} have confirmed this.
From our spin polarized computations having a monovacancy in an 8$\times$8 super cell of graphene, the band structure shows a metallic character and a magnetic moment of $1.15$~$\mu_\mathrm{B}$ (Fig~\ref{fig:band-dos}a) with a magnetization energy of 90~meV.
The non-integer magnetic moment and metallization is mainly because of the periodic boundary conditions, where the corresponding defect density is 0.0078.
Our focus in this study is not the limiting cases but the main features of the defects, therefore we do not consider very large super cells or hybrid functionals in this study.

\begin{figure}[t]
	\centering
	\includegraphics[scale=0.25]{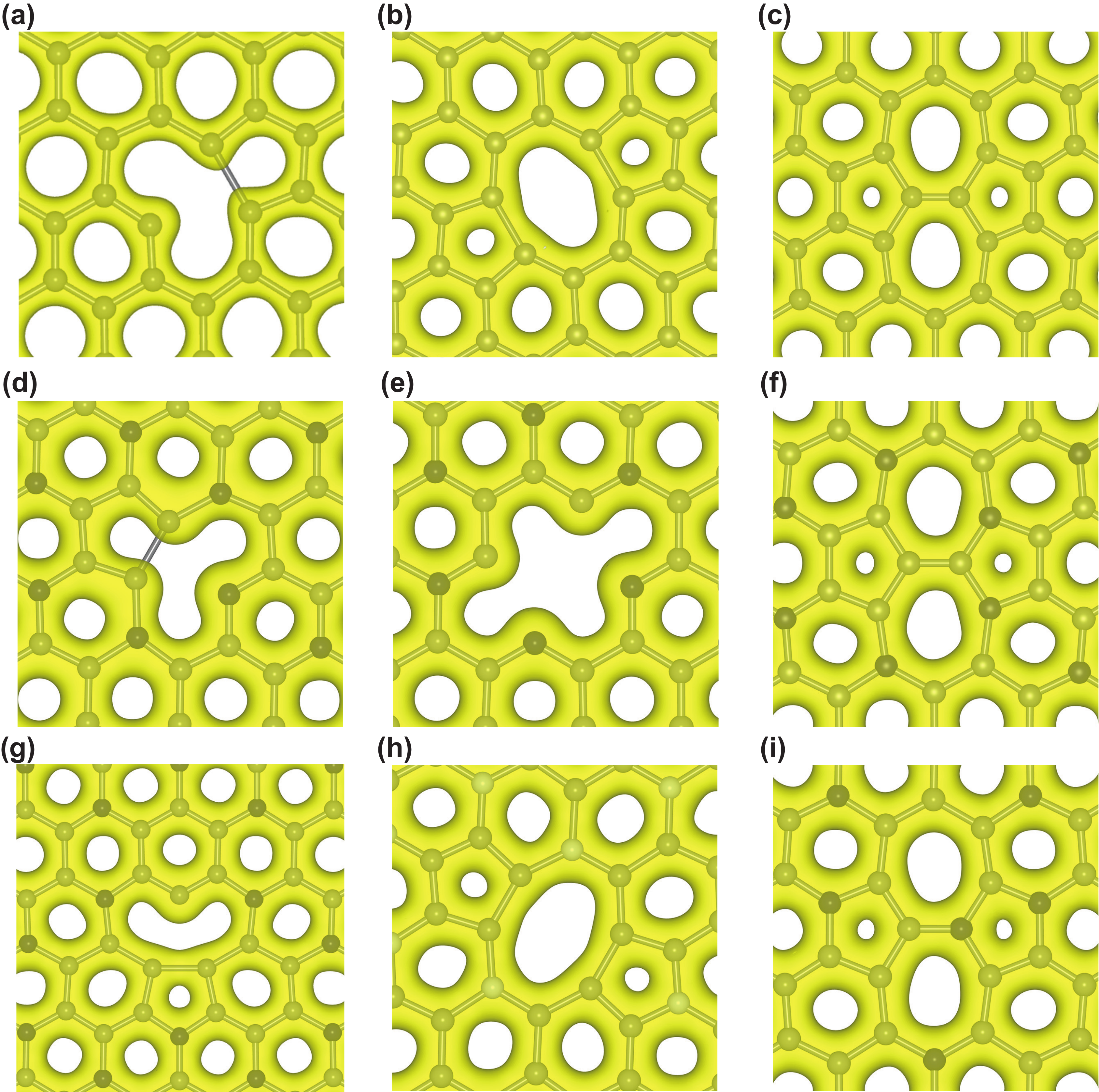}
	\caption{\label{fig:chgu+d} Charge densities of defective graphene and C$_{3}$N structures. (a) gr-MV (b) gr-DV (c) gr-SW (d) C$_{3}$N MV(C) (e) C$_{3}$N DV-C (f) C$_{3}$N Stone-Wales-CC (g) C$_{3}$N MV(N) (h) C$_{3}$N DV(CN) (i) C$_{3}$N Stone-Wales-CN, the isovalue for the charge density plots is 0.075 e/au$^3$. ($\rho=\rho_{\uparrow}+\rho_{\downarrow}$)} 
\end{figure}

In contrast to their similarities in geometrical reconstructions, the monovacancies in \CN have distinctive features from gr-MV. 
Unlike in graphene, removing an atom  (or pairs of atoms for the cases of divacancies) from \CN alters the stoichiometry of the super cell.
This is why, not only \CN-MV but also \CN-DV structures exhibit metallic character, namely partially filled bands, even when all the atoms are three-coordinated.
Low-coordinated atoms give rise to  a range of magnetic moments.
The magnetic moment $m$ and magnetic stabilization energy $\Delta E_{mag}$ values are given in Table~\ref{tab:formation}.

In \CN-MV(C) the low-coordinated atom is nitrogen and the newly formed pentagon consists of carbon atoms, whereas in \CN-MV(N) a carbon atom is low-coordinated and the pentagon includes a nitrogen atom.
The resulting magnetic moments are 0.99~$\muB$ and 0.29~$\muB$ for C and N vacancies, respectively. 
The difference in magnetization is also observed in the spin density difference plots (Fig.~\ref{fig:chgu-d}), where $\Delta s=\rho_\uparrow-\rho_\downarrow$ are plotted.
In \CN-MV(C), $\Delta s$ is localized on two carbon atoms, while it is located on the two-coordinated carbon atom for \CN-MV(N).
In gr-MV, all atoms around the vacancy contribute to $\Delta s$.
We also note that there is higher charge density ($\rho=\rho_\uparrow+\rho_\downarrow$) around the newly formed carbon-carbon bond for \CN-MV(N). (Fig.~\ref{fig:chgu+d})

\begin{figure}[t]
	\centering
	\includegraphics[scale=0.30]{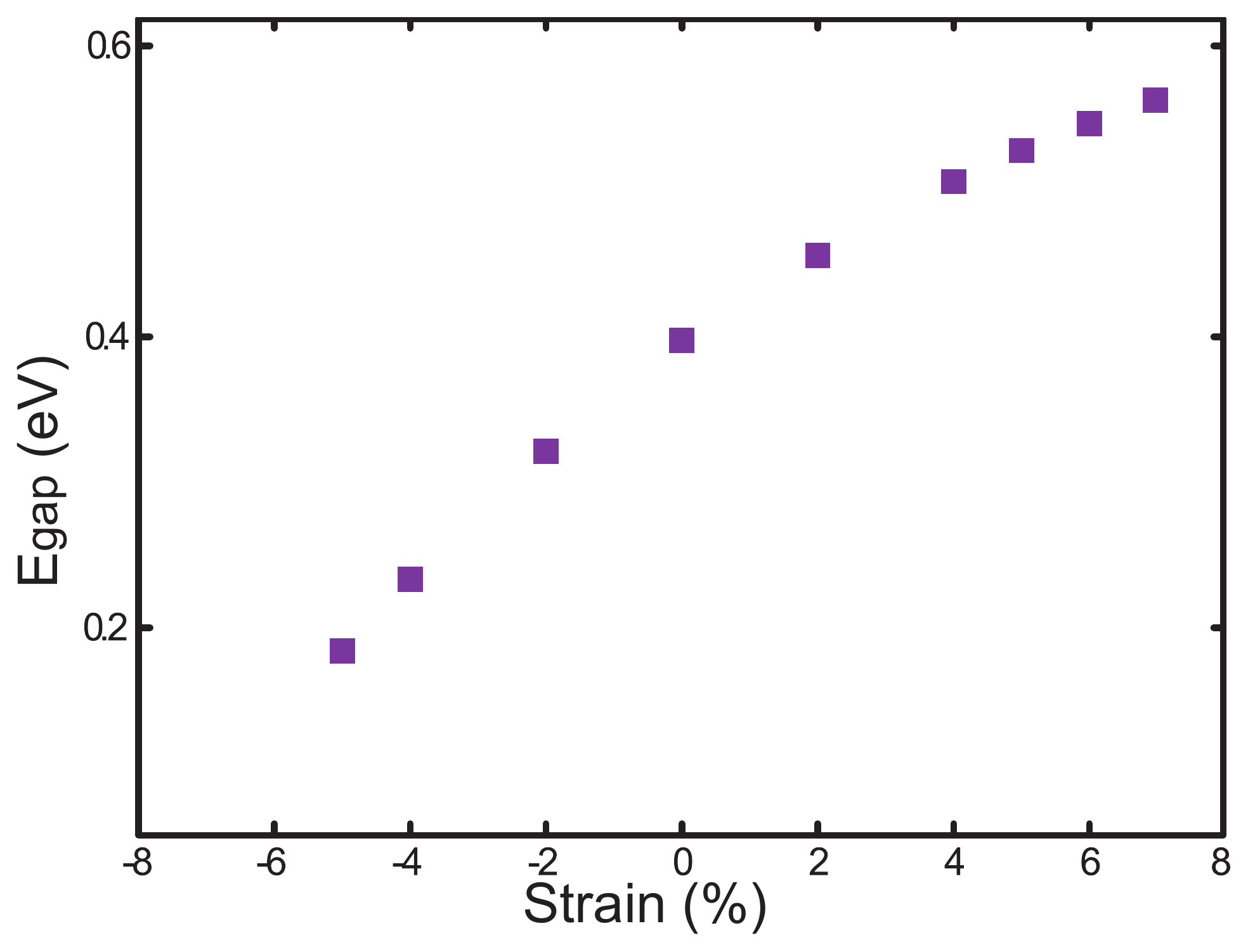}
	\caption{\label{fig:straingraph}  Variation of the band gap of pristine C$_3$N structure as a function of applied biaxial strain. }
\end{figure}

The \CN-DV(CC) structure has four two-coordinated atoms (two carbon and two nitrogen) and a magnetic moment of 1.23~$\muB$.
$\Delta s$ is distributed mainly on the two-coordinated atoms, with carbon atoms being slightly more magnetized (Fig.~\ref{fig:chgu-d})
The \CN-DV(CN) structure has zero magnetic moment, which is interpreted as being due to all atoms being three-coordinated, unlike in \CN-DV(CC) (see Fig.~\ref{fig:yapisal}).
The \CN-SW defects do not alter the stoichiometry and all atoms are three-coordinated, therefore these structures have zero magnetic moment and clear band gaps around the Fermi energy. (see Fig.~\ref{fig:band-dos})
One should also note that, for those structures with a finite magnetic moment flat bands are observed in the electronic structure, which are absent in non-magnetic structures.
For instance, in \CN-MV(C) two spin-up flat bands appear around -0.2~eV and -1.2~eV and two spin-down flat bands appear around -0.75~eV and 1.2~eV, which are associated with the dangling bonds.

\subsection{The effects of strain}

\begin{table}[t]
	\centering
	\caption{\label{tab:strain}Change of magnetic moments of C$_3$N according to applied tensile and compressive biaxial percent strains.}
	\begin{tabular}{ccccccc}
		\hline\hline
		Strain & MV(C) & MV(N) & DV(CC) & DV(CN) & SW(CC) & SW(CN) \\ 
		\hline 
		-4.0 & 0.00 & 0.00 & 0.00 & 0.00 & 0.00 & 0.00 \\ 
		-2.0 & 0.00 & 0.00 & 0.82 & 0.00 & 0.00 & 0.00 \\ 
		0.0 & 0.99 & 0.29 & 1.23 & 0.00 & 0.00 & 0.00 \\ 
		2.0 & 1.07 & 0.30 & 1.23 & 0.00 & 0.00 & 0.00 \\ 
		4.0 & 1.05 & 0.30 & 1.02 & 0.00 & 0.00 & 0.00 \\ 
		\hline\hline
	\end{tabular}  
\end{table}

In order to investigate the robustness of the structures and their properties, we
apply tensile and compressive biaxial strain to defective \CN structures. 
The magnetic moments are given in Table \ref{tab:strain} for strain between $-0.04$ and $0.04$. With increasing compressive biaxial strain, total magnetic moments decrease and finally vanishes. For increasing the tensile strain, the C$_3$N-MV(C), C$_3$N-MV(N), and C$_3$N-DV(CC) obey the same trend. For C$_3$N-DV(CN) and C$_3$N-SW defects, applying tensile and compressive strain do not induce any magnetic moments within the tabulated range of strains. 
When tensile strain is increased even further, a finite magnetic moment of 0.36~$\muB$ is observed for \CN-DV(CN), which is associated with the broken bonds at the defect site.

We have also computed the variation of band gap of pristine C$_3$N under  biaxial  strain. As it is shown in  Fig.~\ref{fig:straingraph}, the energy band gap increases with increasing tensile strain, whereas the compressive strain decreases it down to 0.2~eV. We note that there is no indirect to direct band gap transition under the tensile or compressive biaxial strain.

In order to investigate the stability of defective structures under strain, we perform AIMD for selected structures, namely \CN-DV(CC) and \CN-MV(C). We nothe that both structures have finite magnetic moments. Applying 3.0\% tensile strain, we perform AIMD simulations at $500 K$. We find that both structures are stable during the entire simulation of 10000 steps.

\section{Conclusion}
Structural, electronic and magnetic properties of different point defects such as monovacany, divacancy and Stone-Wales defects on C$_3$N monolayer are investigated by using density functional theory. The results are compared and contrasted with similar defect structures in graphene.
It is shown that some defect types (\CN-MV(C), \CN-MV(N) and \CN-DV(CC)) give rise to magnetization, whereas spin degeneracy is not broken in \CN-DV(CN), \CN-SW(CC) and \CN-SW(CN) defects. 
The magnetization is directly related to lattice reconstructions and their values vary with applied strain. 
Vacancies in \CN could be interesting for spintronics applications and could have advantages over graphene due to its finite band gap.
It should be possible to create defects in \CN using AC-TEM at around 80~kV like in graphene or at smaller energies because defect formation and cohesive energies are smaller than in graphene.

\section{Acknowledgements}
We  acknowledge support from Scientific and Technological Research Council   of   Turkey (TÜB\.{I}TAK)  Grant  No. 117F480. Part of the computations  are  carried out at TÜB\.{I}TAK-ULAKB\.{I}M High Performance and Grid Computing Center.

\section{Data Availability Statement}
The data that support the findings of this study are available from the corresponding author upon reasonable request.

%

\end{document}